\documentclass{PoS}

\title{The adjoint potential in the pseudoparticle approach: string breaking and Casimir scaling}

\ShortTitle{The adjoint potential in the pseudoparticle approach: string breaking and Casimir scaling}

\author{Christian Szasz \\
        University~of~Erlangen-N{\"u}rnberg, Institute~for~Theoretical~Physics~III, Staudtstra{\ss}e~7, 91058~Erlangen, Germany \\
        E-mail: \email{christian.szasz@theorie3.physik.uni-erlangen.de}}

\author{\speaker{Marc Wagner} \\
        Humboldt~University~Berlin, Department~of~Physics, Newtonstra{\ss}e~15, 12489~Berlin, Germany \\
        E-mail: \email{mcwagner@physik.hu-berlin.de}}

\abstract{We perform a detailed study of the adjoint static potential in the pseudoparticle approach, which is a model for SU(2) Yang-Mills theory. We find agreement with the Casimir scaling hypothesis and there is clear evidence for string breaking. At the same time the potential in the fundamental representation is linear for large separations. Our results are in qualitative agreement with results from lattice computations.}

\FullConference{8th Conference Quark Confinement and the Hadron Spectrum \\
		 September 1-6, 2008 \\
		 Mainz, Germany}

\newcommand{\gtapprox}{\raisebox{-0.5ex}{$\,\stackrel{>}{\scriptstyle\sim}\,$}}

\begin{document}


\section{Introduction}

A common approach to gain a better understanding of Yang-Mills theory, in particular the mechanism of confinement, is to restrict the full path integral to a small subset of gauge field configurations, which are supposed to be of physical importance. Examples are instanton gas and liquid models (cf.\ \cite{Schafer:1996wv} and references therein), ensembles of regular gauge instantons and merons \cite{Lenz:2003jp,Negele:2004hs,Lenz:2007st}, the pseudoparticle approach \cite{Wagner:2005vs,Wagner:2006qn,Wagner:2006du,Szasz:2008qk}, calorons with non-trivial holonomy \cite{Gerhold:2006sk,Gerhold:2006kw}, and models based on center vortices (cf.\ e.g.\ \cite{Faber:1997rp,Engelhardt:1999wr,Engelhardt:2003wm,Rafibakhsh:2007sh,Faber:2008}).

In this paper we apply the pseudoparticle approach to SU(2) Yang-Mills theory and perform a detailed study of the static potential for various representations.


\section{The pseudoparticle approach in SU(2) Yang-Mills theory}

The basic idea of the pseudoparticle approach is to approximate the Yang-Mills path integral
\begin{eqnarray}
\label{EQN001} \Big\langle \mathcal{O} \Big\rangle \ \ = \ \ \frac{1}{Z} \int DA \, \mathcal{O}[A] e^{-S[A]} \quad , \quad S[A] \ \ = \ \ \frac{1}{4 g^2} \int d^4x \, F_{\mu \nu}^a F_{\mu \nu}^a ,
\end{eqnarray}
where $F_{\mu \nu}^a = \partial_\mu A_\nu^a - \partial_\nu A_\mu^a + \epsilon^{a b c} A_\mu^b A_\nu^c$, with a small number of physically relevant degrees of freedom. To this end, the integration over all gauge field configurations in (\ref{EQN001}) is restricted to a small subset, which can be written as a linear superposition of a fixed number of pseudoparticles\footnote{In this paper the term pseudoparticle refers to any gauge field configuration $a_\mu^a$, which is localized in space and in time, not only to solutions of the classical Yang-Mills equations of motion.}:
\begin{eqnarray}
\label{EQN002} A_\mu^a(x) \ \ = \ \ \sum_j \mathcal{A}(j) \mathcal{C}^{a b}(j) a_\mu^b(x-z(j)) ,
\end{eqnarray}
where $j$ is the pseudoparticle index and $\mathcal{A}(j) \in \mathbb{R}$, $\mathcal{C}^{a b}(j) \in \textrm{SO(3)}$ and $z(j) \in \mathbb{R}^4$ are the amplitude, the color orientation and the position of the $j$-th pseudoparticle respectively. The functional integration over all gauge field configurations is defined as the integration over pseudoparticle amplitudes and color orientations:
\begin{eqnarray}
\int DA \, \ldots \quad = \quad \int \left(\prod_j d\mathcal{A}(j) \, d\mathcal{C}(j)\right) \ldots
\end{eqnarray}

For the results presented in this work we have used $625$ ``long range pseudoparticles'', which fall off as $1 / \textrm{distance}$, inside a hypercubic spacetime region (for details regarding this setup cf.\ \cite{Szasz:2008qk}):
\begin{eqnarray}
a_{\mu,\textrm{\scriptsize inst.}}^a(x) \ \ = \ \ \frac{\eta_{\mu \nu}^a x_\nu}{x^2 + \lambda^2} \quad , \quad a_{\mu,\textrm{\scriptsize antiinst.}}^a(x) \ \ = \ \ \frac{\bar{\eta}_{\mu \nu}^a x_\nu}{x^2 + \lambda^2} \quad , \quad a_{\mu,\textrm{\scriptsize akyron}}^a(x) \ \ = \ \ \frac{\delta^{a 1} x_\mu}{x^2 + \lambda^2} .
\end{eqnarray}
The first two types generate transverse gauge field components and are similar to regular gauge instantons and antiinstantons, while the third type, the so-called akyron \cite{Wagner:2006qn}, is responsible for longitudinal gauge field components. We would like to stress that gauge field configurations (\ref{EQN002}) are in general not even close to solutions of the classical Yang-Mills equations of motion, i.e.\ the pseudoparticle approach is not a semiclassical model. The idea is rather to approximate physically relevant gauge field configurations by a small number of degrees of freedom.


\section{Casimir scaling and adjoint string breaking}

In the following the potential associated with a pair of static color charges $\phi^{(J)}$ and $(\phi^{(J)})^\dagger$ in spin-$J$-representation at separation $R$ is denoted by $V^{(J)}(R)$. In pure Yang-Mills theory there is no string breaking, when the charges are in the fundamental representation ($J = 1/2$). For charges in the adjoint representation ($J = 1$) the situation is different: gluons are able to screen such charges and the connecting gauge string is expected to break, when the charges are separated adiabatically  beyond a certain distance; a pair of essentially non-interacting gluelumps is formed.

The starting point to extract the static potential in spin-$J$-representation are ``string trial states''
\begin{eqnarray}
\label{EQN003} S^{(J)}(\mathbf{x},\mathbf{y}) | \Omega \rangle \ \ = \ \ (\phi^{(J)}(\mathbf{x}))^\dagger U^{(J)}(\mathbf{x};\mathbf{y}) \phi^{(J)}(\mathbf{y}) | \Omega \rangle \quad , \quad |\mathbf{x}-\mathbf{y}| \ \ = \ \ R ,
\end{eqnarray}
where $U^{(J)}$ denotes a spatial parallel transporter. We compute temporal correlation functions
\begin{eqnarray}
\mathcal{C}_\textrm{\scriptsize string}^{(J)}(T) \ \ = \ \ \langle \Omega | \Big(S^{(J)}(\mathbf{x},\mathbf{y},T)\Big)^\dagger S^{(J)}(\mathbf{x},\mathbf{y},0) | \Omega \rangle \ \ \propto \ \ \Big\langle W_{(R,T)}^{(J)} \Big\rangle
\end{eqnarray}
and determine the corresponding potential values from their exponential fall-off (for details cf.\ \cite{Szasz:2008qk}).

\begin{figure}[b]
\begin{center}
\input{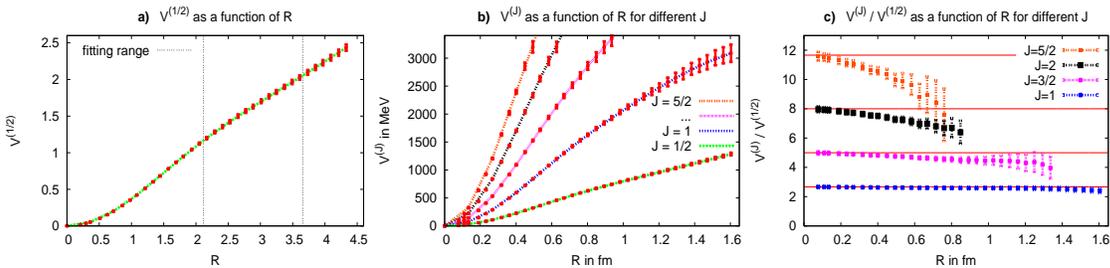}
\caption{\label{FIG001}
\textbf{a)}~The fundamental static potential $V^{(1/2)}$ as a function of the separation $R$.
\textbf{b)}~``Pure Wilson loop static potentials'' $V^{(J)}$ for different representations as functions of the separation $R$.
\textbf{c)}~Ratios $V^{(J)} / V^{(1/2)}$ as functions of the separation $R$ compared to the Casimir scaling expectation.
}
\end{center}
\end{figure}

The numerical result for the fundamental potential is shown in Figure~\ref{FIG001}a (here and in the following we have used the value $g = 12.5$ for the coupling constant). It is linear for large separations, i.e.\ there is confinement. We set the physical scale by fitting $V^{(1/2)}(R) = V_0 + \sigma R$ and by identifying the string tension $\sigma$ with $\sigma_\textrm{\scriptsize physical} = 4.2 / \textrm{fm}^2$. This amounts to a spacetime region of extension $L^4 = (1.85 \, \textrm{fm})^4$.

Numerical results for higher representation potentials ($J=1,\ldots,5/2$) are shown in Figure~\ref{FIG001}b. According to the Casimir scaling hypothesis these potentials are supposed to fulfill
\begin{eqnarray}
V^{(1/2)}(R) \ \ \approx \ \ \frac{V^{(1)}(R)}{8/3} \ \ \approx \ \ \frac{V^{(3/2)}(R)}{5} \ \ \approx \ \ \frac{V^{(2)}(R)}{8} \ \ \approx \ \ \frac{V^{(5/2)}(R)}{35/3}
\end{eqnarray}
for intermediate separations. Figure~\ref{FIG001}c shows that this is the case for the adjoint potential, while there are slight deviations for $J \geq 3/2$. This is in agreement with what has been observed in 4d SU(2) lattice gauge theory \cite{Piccioni:2005un}.

\begin{figure}[b]
\begin{center}
\input{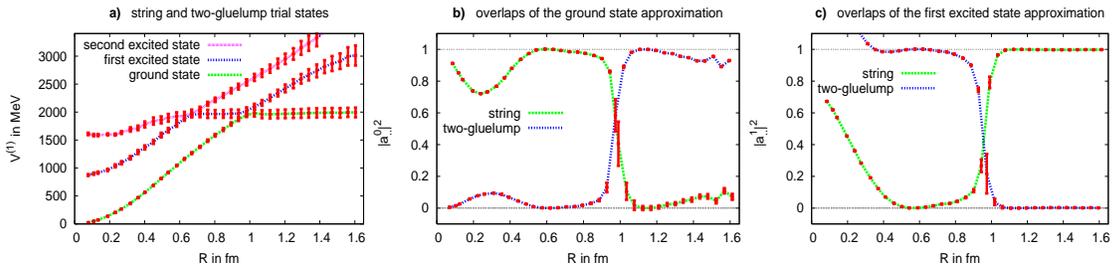}
\caption{\label{FIG002}
\textbf{a)}~The adjoint static potential $V^{(1)}$ and its first two excitations as functions of the separation $R$.
\textbf{b)}~Overlaps of the ground state approximation to the trial states as functions of the separation $R$.
\textbf{c)}~Overlaps of the first excited state approximation to the trial states as functions of the separation $R$.
}
\end{center}
\end{figure}

Note that there is no sign of string breaking for the adjoint potential even for separations $R \gtapprox 1.6 \, \textrm{fm}$. This is, because string trial states have poor overlap to the ground state, which is expected to resemble a two gluelump state. The solution to overcome this problem is to use a whole set of trial states containing not only string trial states (\ref{EQN003}), but also ``two-gluelump trial states''
\begin{eqnarray}
\sum_{j = x,y,z} G_j(\mathbf{x}) G_j(\mathbf{y}) | \Omega \rangle \quad , \quad G_j(\mathbf{x}) \ \ = \ \ \textrm{Tr}\Big(\phi^{(1)}(\mathbf{x}) B_j(\mathbf{x})\Big) \quad , \quad |\mathbf{x}-\mathbf{y}| \ \ = \ \ R .
\end{eqnarray}
We extract the adjoint potential from the corresponding correlation matrices by solving a generalized eigenvalue problem and by computing effective masses  (for details cf.\ \cite{Szasz:2008qk}). Results are shown in Figure~\ref{FIG002}a. The potential saturates at around two times the magnetic gluelump mass (which is $\approx 1000 \, \textrm{MeV}$ at $g = 12.5$ in this regularization \cite{Szasz:2008qk}) at separation $R_\textrm{\scriptsize sb} \approx 1.0 \, \textrm{fm}$. This string breaking distance as well as the observed level ordering (the first excited state is an excited string state for small separations, then becomes a two gluelump state and finally a string state again, etc.) is in agreement with results from lattice computations \cite{Jorysz:1987qj,deForcrand:1999kr}.

To investigate, whether the gluonic string really breaks, when two static charges are separated adiabatically, we perform a mixing analysis. During the computation of effective masses we obtain approximations of the ground state and the first excited state,
\begin{eqnarray}
| 0 \rangle \ \ \approx \ \ a_\textrm{\scriptsize string}^0 | \textrm{string} \rangle + a_\textrm{\scriptsize 2g-lump}^0 | \textrm{2g-lump} \rangle \quad , \quad | 1 \rangle \ \ \approx \ \ a_\textrm{\scriptsize string}^1 | \textrm{string} \rangle + a_\textrm{\scriptsize 2g-lump}^1 | \textrm{2g-lump} \rangle ,
\end{eqnarray}
where $| \textrm{string} \rangle$ and $| \textrm{2g-lump} \rangle$ are normalized trial states. The overlaps $|a_{\ldots}^j|^2$ are shown as functions of the separation in Figure~\ref{FIG002}b and \ref{FIG002}c. The transition between string and two-gluelump states is rapid but smooth indicating that string breaking is present in the pseudoparticle approach.


\section{Conclusions and outlook}

We have computed static potentials for various representations within the pseudoparticle approach. While the fundamental static potential is linear for large separations, we clearly observe string breaking for the adjoint representation. Both the string breaking distance $R_\textrm{\scriptsize sb} \approx 1.0 \, \textrm{fm}$ and the level ordering are in agreement with lattice results, and a mixing analysis indicates a rapid, but smooth transition between a string and a two gluelump state, when two static charges are separated adiabatically. Moreover, higher representation potentials exhibit Casimir scaling. We conclude that the pseudoparticle approach is a model, which is able to reproduce many essential features of SU(2) Yang-Mills theory.

Currently our efforts are focused on applying the pseudoparticle approach to fermionic theories. First steps in this direction have been successful \cite{Wagner:2007he,Wagner:2007av}. Now we intend to consider QCD, where a cheap computation of exact all-to-all propagators should be possible due to the small number of degrees of freedom involved. Another appealing possibility is an application to supersymmetric theories, where an exact realization of supersymmetry might be possible due to translational invariance present in pseudoparticle ensembles.


\begin{acknowledgments}

MW would like to thank M.~Faber, J.~Greensite and M.~Polikarpov for the invitation to this conference. Moreover, we acknowledge useful conversations with M.~Ammon, G.~Bali, P.~de~Forcrand, H.~Hofmann, E.-M.~Ilgenfritz, F.~Lenz and M.~M\"uller-Preussker.

\end{acknowledgments}



\end{document}